\documentclass[10pt,a4paper,twoside]{article}
\usepackage{epsfig}
\usepackage{natbib}
\usepackage{baltlat5}
\usepackage{wrapfig}
\def\la{\mathrel{\mathchoice {\vcenter{\offinterlineskip\halign{\hfil
$\displaystyle##$\hfil\cr<\cr\sim\cr}}}
{\vcenter{\offinterlineskip\halign{\hfil$\textstyle##$\hfil\cr
<\cr\sim\cr}}}
{\vcenter{\offinterlineskip\halign{\hfil$\scriptstyle##$\hfil\cr
<\cr\sim\cr}}}
{\vcenter{\offinterlineskip\halign{\hfil$\scriptscriptstyle##$\hfil\cr
<\cr\sim\cr}}}}}
\def\ga{\mathrel{\mathchoice {\vcenter{\offinterlineskip\halign{\hfil
$\displaystyle##$\hfil\cr>\cr\sim\cr}}}
{\vcenter{\offinterlineskip\halign{\hfil$\textstyle##$\hfil\cr
>\cr\sim\cr}}}
{\vcenter{\offinterlineskip\halign{\hfil$\scriptstyle##$\hfil\cr
>\cr\sim\cr}}}
{\vcenter{\offinterlineskip\halign{\hfil$\scriptscriptstyle##$\hfil\cr
>\cr\sim\cr}}}}}

\begin{document}
\ \
\vspace{-0.5mm}

\setcounter{page}{1}
\vspace{-2mm}

\titlehead{Baltic Astronomy, vol.\ts yy, xxx--xxx, 2006.}

\titleb{BINARY LIFE AFTER THE AGB - TOWARDS A UNIFIED \\ PICTURE}

\begin{authorl}
\authorb{A. Frankowski}{} and
\authorb{A. Jorissen}{}
\end{authorl}

\moveright-3.2mm
\vbox{
\begin{addressl}
\addressb{}{Institut d'Astronomie et d'Astrophysique, Universit\'e
  Libre de Bruxelles, Boulevard du Triomphe CP 226, B-1050 Bruxelles, Belgium}
\end{addressl}
}

\submitb{Received 2006 xx; revised 2006 xx}

\begin{summary}
A unified evolutionary scheme that includes post-AGB systems, barium stars,
symbiotics, and related systems, explaining their similarites as well as
differences. Can we construct it?
We compare these various classes of objects in order to construct a
consistent
picture. Special attention is given to the comparison of the
barium pollution and symbiotic phenomena. Finally, we outline a
`transient torus' evolutionary scenario that makes use of the various
observational and theoretical hints and aims at explaining the observed
characteristics of the relevant systems.
\end{summary}


\begin{keywords}
stars: AGB and post-AGB, stars: binaries
\end{keywords}

\resthead{Binary life after the AGB}{A. Frankowski, A. Jorissen}

\sectionb{1}{AFTER-AGB BINARIES}
\label{Sect:intro}

The term ``after-AGB binaries'' will be used in this paper to refer
to binary systems in which at
least {\em one of the components} has gone through the AGB phase, as
distinguished from the term ``post-AGB'', commonly used to denote the
short transition phase between AGB and PN core (CSPN)
stages of (single or binary) stellar evolution. In many of such
after-AGB systems, the mass transfer from an AGB star has left its mark
on the companion, enhancing its abundances with the products of the AGB
nucleosynthesis, most remarkably C, F, and s-process elements
(see Habing \& Olofsson 2003 for a recent review).
This pollution will manifest itself later in the companion's evolution.
An exemplary case of after-AGB systems are barium stars:
G-K type giants remarkable for their overabundances of Ba
(McClure et~al.\ 1980). Related families include Abell-35 subclass of PNe
(Bond et~al.\ 1993),
barium dwarfs
(including the so called WIRRing stars, Jeffries \& Stevens 1996),
subgiant and giant CH stars, extrinsic S stars and d'-type yellow
symbiotics.
But not all of the after-AGBs need to be s-process rich. The post-AGB
binaries are an interesting case, as they are all by definition after-AGBs:
some of them do exhibit s-process enhancement while others do not
(Van Winckel 2003, also this volume).
Red s-type symbiotic stars (SyS) with massive white dwarf companions
($M_{\rm WD}>0.5 M_{\odot}$), another member of the after-AGB group,
also do not exhibit s-process enhancement (Jorissen 2003a).
Not much in this respect can be said about most
binary CSPNe, as the unevolved companion is usually too
faint to be seen.
Finally, some of the cataclysmic variables (CV) with massive
white dwarfs should also belong to the after-AGB family.

\vbox{
\vspace{-3mm}
\centerline{\psfig{figure=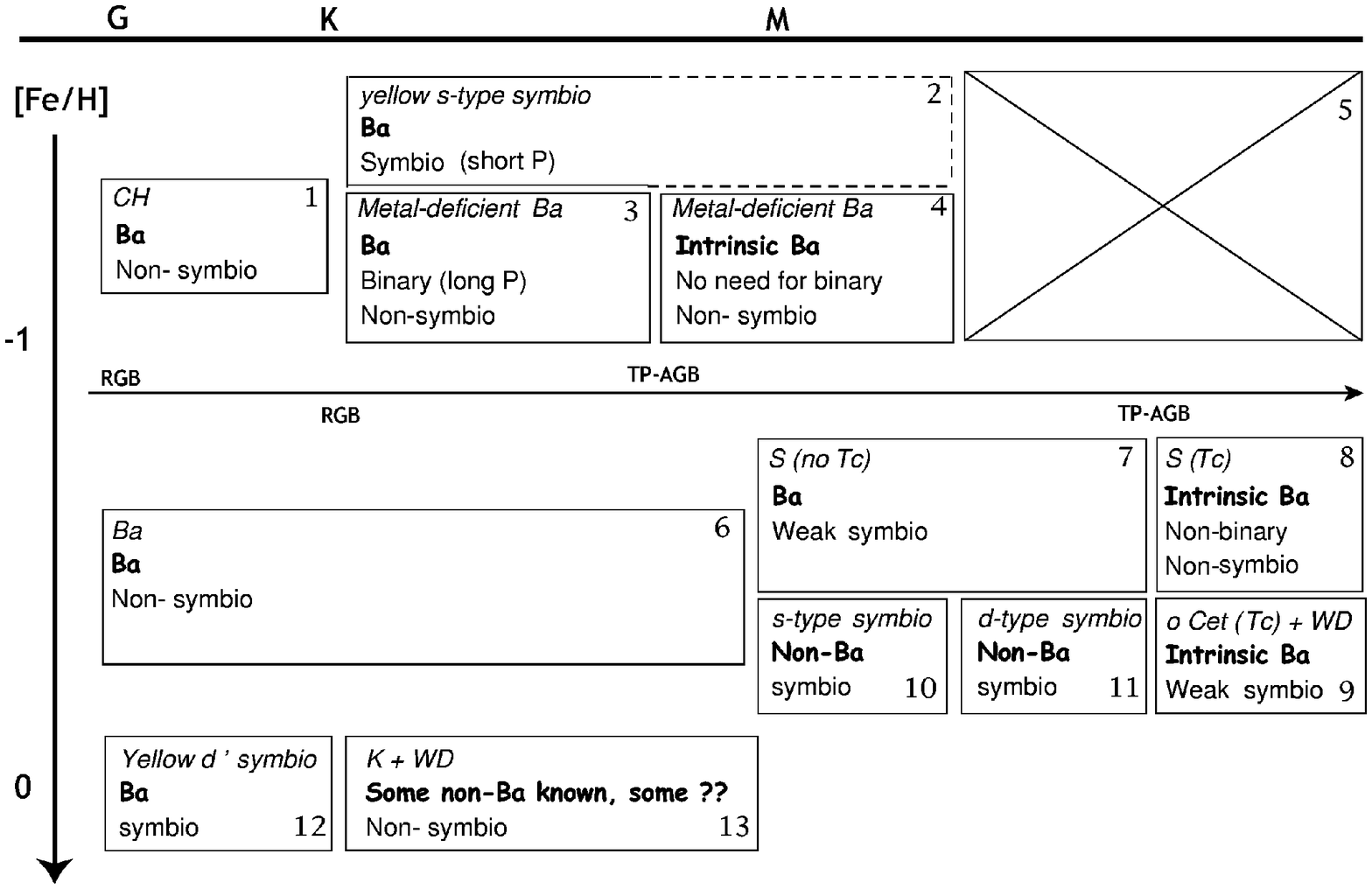,width=120truemm,clip=}}
\label{Fig:links}
\vspace{-3mm}
\captionb{1}{
The various kinds of SyS and of peculiar red giants, in a plane
spectral-type -- metallicity. Within each box, the first line (slanted
font) lists the stellar family, the second line (bold font) indicates
whether or not stars from that family are enriched in s-process
elements (either from internal nucleosynthesis -- ``Intrinsic Ba'' --
or from mass transfer across a binary system -- ``Ba'' standing for
Extrinsic Ba). The last (two) line(s) (normal font) provide(s) the
binary properties of that family: binary (short or long periods) or 
non-binary, SyS or non-SyS.   
Each box has been assigned a number, for easy reference in the text. 
The line with an arrow in the middle of the figure is to indicate that the
TP-AGB phase involves different spectral types
at low- and near-solar metallicity.
}
}
\vspace{0mm}

\sectionb{2}{LINKS BETWEEN SYS AND PECULIAR RED GIANTS}
\label{Sect:links}

Peculiar red giants are a characteristic part of the after-AGB family
and are in many respects closely related to symbiotic stars.
The links between them have been
reviewed previously (Jorissen 2003a, Jorissen et~al.\ 2005); 
here we provide an updated discussion
based on Fig.~1 which displays the various types of
SyS and peculiar red giants in a metallicity --
spectral-type plane.

The vertical axis corresponds to metallicity which impacts (i) the 
taxonomy of the classes (CH giants for
instance -- box~1 -- are the halo-equivalent of the disk barium stars
-- box~6); (ii) the efficiency of heavy-element synthesis
(Clayton 1988), and (iii)  the location of evolutionary tracks in
the HR diagram (hence the correspondence between spectral type and
evolutionary status, like the onset of TP-AGB, will depend on
metallicity).
Fig.~1 therefore considers three different metallicity
ranges: (i) [Fe/H]$<-1$, corresponding to the halo population; (ii)
$-1 \la [Fe/H] \la 0$, or disk metallicity; (iii) [Fe/H]$\ga 0$,
solar and super-solar metallicities. 
The horizontal axis in Fig.~1 displays spectral type,
which roughly corresponds  to an evolutionary 
sequence at a given metallicity, passing from the red giant branch
(RGB) to the thermally-pulsing AGB (TP-AGB) phases (between these two
phases is the core He-burning phase which is hardly distinguishable from
the lower RGB; CH giants probably belong to that phase). 
Symbiotic activity is expected in the middle of
this sequence, because (i) at the left end, stars (like CH) are not
luminous enough to experience a mass loss sufficient to power
symbiotic activity; (ii) at the right end, the stars with the barium
syndrome need not be binaries 
(above the TP-AGB luminosity threshold,
heavy-elements are synthesized in the stellar interior and dredge-up
to the surface, so that
``intrinsic Ba'' (or S) stars occupy the rightmost boxes -- 4 and 8 -- of
Fig.~1), and hence need not exhibit
any symbiotic activity. Those evolved giants which {\em are} members of
binary systems (like Mira Ceti -- box~9) 
will of course exhibit symbiotic activity.
It is noteworthy that late M giants are inexistent in a
halo population (hence the crossed box~5), because evolutionary tracks
are bluer as compared to higher metallicities. Examples of such very
evolved (relatively warm) 
stars in a halo population (box~4) include CS~30322-023
(Masseron et~al.\ 2006) and V~Ari (Van Eck et~al.\ 2003).

\subsectionb{2.1}{Halo}

In the halo population, the most interesting issue
is to understand the origin of the difference
between yellow s-type SyS (box~2) and metal-deficient Ba stars
(box~3): why do the latter not exhibit any
symbiotic activity? The reason seems to reside in a difference
in their orbital period distributions: yellow s-type SyS
represent the short-period tail of the distribution
(Fig.~2), where for a given mass-loss rate accretion
will be more efficient, and hence will trigger symbiotic activity.

\subsectionb{2.2}{Intermediate metallicities}

\begin{wrapfigure}[22]{r}[0pt]{68mm}
\vbox{
\vspace{-5mm}
\centerline{\psfig{figure=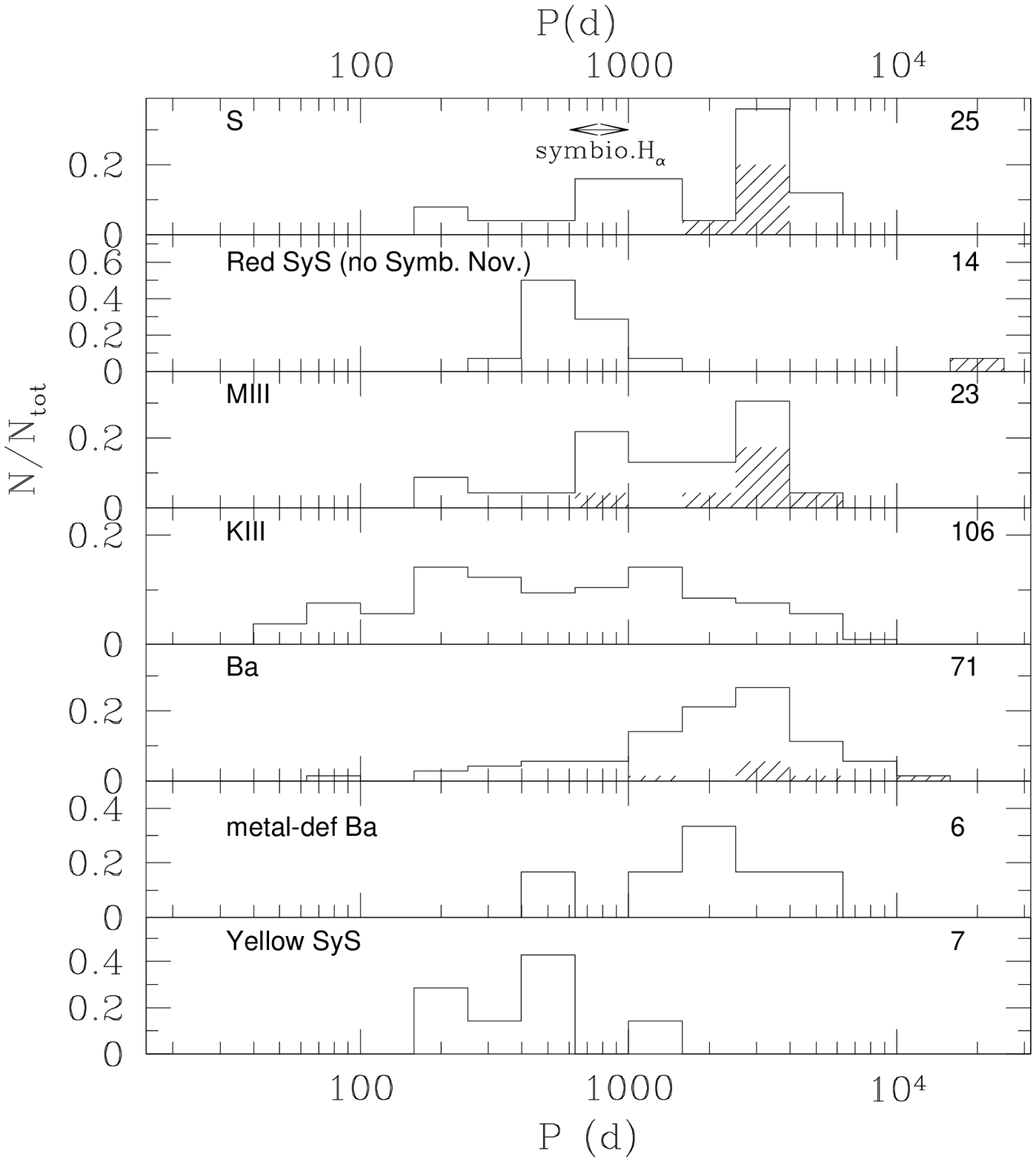,width=72mm,clip=}}
\captionb{2}{\label{Fig:P_histo}
Histograms of the orbital period distribution for various families of
binary red giants. From Jorissen et~al.\ (2005).
}
}
\end{wrapfigure}

In the intermediate-metalli\-ci\-ty regime, the situation is quite clear:
as the star evolves to the right along the spectral sequence, 
its luminosity (hence mass loss) increases, and for binary stars along
that sequence, the symbiotic character will become stronger.    
The real issue here is to understand the origin of the difference
between boxes 7 and 10/11: why are red s- (and d-)type SyS never
exhibiting the barium syndrome, despite very similar locations in the
HR diagram and similar orbital-period distributions
(Fig.~2)? In former discussions of this issue 
(e.g.\ Jorissen 2003a), it had been suggested that binary stars
with the barium syndrome (box~7) and SyS without it
(box~10) differ in their metallicities.
It is known that at high metallicities heavy-element
synthesis is less efficient (Clayton 1988). However, there is so far
{\em no evidence for red symbiotic stars being on
average more metal-rich than barium or extrinsic S stars.}
Schild et~al.\ (1992) and Schmidt \& Miko\l ajewska (2003) have compared
the carbon abundances of SyS and normal giants, and found
absolutely no difference, thus confirming at the same time the
  absence of any signature of internal
  nucleosynthesis and dredge-ups, or of pollution through mass transfer.   

Other solutions to this puzzle may be suggested, such as 
(i) s- and d-type SyS are not intrinsic barium stars, because
they are not TP-AGB stars;
(ii) neither are they extrinsic barium stars, because their
companion never went through the TP-AGB phase, either because it is a
He WD or because it is a main sequence star. 
Regarding item~(i), it is very likely indeed that red s-type
SyS are not TP-AGB stars, since they rather involve early M
giants. The situation is less clear for d-type SyS, as they involve
Miras which are often claimed to be TP-AGB stars. Nevertheless, 
many Miras do not exhibit signatures of heavy-element
nucleosynthesis (they are not carbon stars and lack lines from the
unstable element Tc; Little et~al.\ 1987). 

Regarding item~(ii), the possibility for hot companions to SyS 
to be He WDs is the most appealing since (a) the 
eccentricities observed for symbiotic systems are much smaller than
those observed in pre-mass-transfer systems (M giants in the
  period range 200 -- 1000~d have eccentricities up to
  0.3; Jorissen et~al.\ 2004), thus suggesting that mass transfer {\em
  has taken place} in these systems; (b) the mass distribution of the
hot components peaks between 0.4 and 0.5~$M_\odot$\ 
(Miko\l ajewska 2003 and this volume), as expected for He WDs.  
Of course, an alternative explanation -- like a main sequence
companion -- needs to be found for 
those SyS companions with masses exceeding 0.5~$M_\odot$\ (T~CrB,
  FG~Ser, FN~Sgr, AR~Pav, V1329~Cyg; Miko\l ajewska 2003). 
Although a main sequence companion is quite unlikely in recurrent or
symbiotic novae like T~CrB and V1329~Cyg, the situation regarding the
nature of symbiotic-star companions for non-nova systems is far from
being settled, as mentioned by Miko\l ajewska (2003) while answering 
a question by one of the
authors at the La Palma symbiotic-star conference: {\em
  the question of whether [the companion to CI~Cyg, Z~And, FN~Sgr] is
  a disk-accreting main-sequence star or a quasi-steady
  hydrogen-burning white dwarf is open so long as we have no good
  theory to distinguish between these possibilities}. 
Indeed, the nature of the companion to CI~Cyg has changed over the
years, from main-sequence accretor 
(Kenyon \& Webbink 1984; Kenyon et~al.\ 1991;
Miko\l ajewska \& Kenyon 1992) to 
hot and luminous stellar source powered by thermonuclear burning 
(Miko\l ajewska 2003)! The same move from main-sequence to
white-dwarf accretor holds true for AR~Pav
(Kenyon \& Webbink 1984,Quiroga et~al.\ 2002).

\subsectionb{2.3}{Solar metallicities}
\label{Sect:solar_Z}

The evolutionary status of the rare set of yellow d' SyS (box~12),
which were all shown to be of solar metallicity, has recently 
been clarified (Jorissen et~al.\ 2005) 
with the realisation that in these systems, the
companion is {\em intrinsically} hot (because it recently 
evolved off the AGB), rather than being powered by accretion or
nuclear burning. Several arguments support this claim: 
(i) d'~SyS host G-type giants whose mass loss is not strong enough to
heat the companion through accretion and/or nuclear burning;  
(ii) the cool dust observed in d' SyS
(Schmid \& Nussbaumer 1993) is a relic from the mass lost by the AGB
star; 
(iii) the optical nebulae observed in d' SyS are most likely 
genuine planetary nebulae (PN) rather than the 
nebulae associated with the ionized wind of the cool component
(Corradi et~al.\ 1999). d'~SyS often appear in PN catalogues.    
AS~201 for instance actually hosts {\em two} nebulae (Schwarz 1991): 
a large fossil planetary nebula detected by direct imaging, and a small
nebula formed in the wind of the current cool component;
(iv) rapid rotation is a common property of the cool components of  d'
SyS (see Table~1 of Jorissen et~al.\ 2005). 
It has likely been caused by spin accretion from the
former AGB wind like in WIRRing systems  
(Jeffries \& Stevens 1996; Jorissen 2003b). The fact that the cool star has
not yet been slowed down by magnetic braking is another indication
that the mass transfer occurred fairly recently (Theuns et~al.\ 1996). 
Corradi \& Schwarz (1997)
obtained 4000~y for the age of the nebula around AS~201, 
and 40000~y for V417~Cen. 

The possible existence, in box~13, of binary systems of nearly solar
metallicity with orbital properties typical of barium systems, but not
exhibiting the barium syndrome, is still controversial, as discussed
by Jorissen (2003b).

\sectionb{3}{ORBITAL CHARACTERISTICS: RIDDLES AND HINTS}
\label{Sect:orbital}

Intense AGB mass loss/transfer is not only important for chemical
abundances; it does also influence the orbital properties of after-AGB
systems.
Four binary evolution processes are usually invoked when describing the
after-AGB systems formation: (i) tidal interactions,
(ii) wind accretion, including tidally enhanced winds
(Companion-Reinforced Attrition Process or CRAP, Eggleton 1986),
(iii) stable Roche-lobe overflow (RLOF), and
(iv) common envelope (CE) evolution.

Alas, current
evolutionary computations fail to reproduce the correct ranges of orbital
periods, eccentricities and s-process enhancement levels
(e.g.\ Pols et~al.\ 2003; Frankowski 2004).
The basic reason for this problem is quite simple, as nicely put by
Iben \& Tutukov (1996):
{\em as a result of CE interaction, initially close systems become closer
and, because of wind mass loss, initially wide systems become wider. [...]
most known symbiotic systems belong to a rare population on the borderline
between initially close and wide binaries.}
The models do not produce eccentric
systems with periods below $\sim2000$ -- 3000\,d and all systems below
$\sim$1000\,d enter a CE and undergo a dramatic orbital shrinkage.

The observed after-AGB systems with intermediate periods (100 -- 2000\,d)
have somehow avoided the catastrophic outcome of a CE, but
the theoretical concepts proposed so far are not satisfactory in explaining
this fact:
(i) the inclusion of tidal forces affects
only the detached evolution and does not improve the final results;
(ii) CRAP does allow for slightly shorter final periods in detached evolution
but still not below 2000\,d and any stronger effect would prevent TP-AGB,
thus impeding s-process;
(iii) stable RLOF occurs only for a narrow range of initial parameters;
(iv) lowered binding energy of the AGB envelope due to the inclusion of
ionisation energy as proposed e.g.~by Han et~al.\ (1994) is problematic
(Harpaz 1998);
(v) CE formalism based on angular-momentum instead of on energy
(Nelemans et~al.\ 2000) is promising, however, for the moment it lacks
physical explanation.

Also puzzling is high eccentricity (up to $e\!=\!0.4$) at periods down
to 300\,d observed among post-AGB binaries, and to a lesser extent Ba
stars and extrinsic S stars.
The most promising explanation here is eccentricity pumping by
a circumbinary disk (Waelkens et~al.\ 1996; Artymowicz et~al.\ 1991).
Another suggestion is periastron mass loss eccentricity pumping
(Soker 2000) but this mechanism can operate only for wide (detached on
the AGB) systems.

We suggest that these conundrums are part of a bigger puzzle together with
the following observational and theoretical hints. First, some of the young
after-AGB objects exhibit combined RS CVn and Ba star properties:
X-rays, H$_\alpha$ emission and fast rotation combined with Ba enhancement
and long orbital periods. The list consists of Ba stars
56 Peg (Frankowski \& Jorissen 2006),
HD 165141 (Jorissen et~al.\ 1996),
d' symbiotics
(Jorissen et~al.\ 2005, see also the discussion in
Sect.~2.3),
WIRRing stars, (Jeffries \& Stevens 1996), 
and Abell-35 CSPNe (Th\'evenin \& Jasniewicz 1997).
They form a strong evidence for fast rotation in young after-AGB systems,
supposedly due to spin accretion from wind
(Jeffries \& Stevens 1996; Jorissen 2003b).
Second, post-AGB systems, the youngest among the after-AGB family, are
known to possess circumbinary disks (Van Winckel 2003). Dusty
circumbinary disks,
tori and bipolar outflows are common among bipolar and ring-like PNe, and
have also been observed in some AGB stars, notably $\pi^1$ Gru
(Sahai 1992)
and V Hya (Knapp et~al.\ 1999). The latter object is also remarkable for
having fast rotation velocity (6-16 km s$^{-1}$) and a long secondary
photometric period ($\sim 6200$\,d,
in addition to the radial pulsation period of 530\,d),
possibly due to a binary companion.
Another notable factor is that dust formation and radiation-driven wind
cause reshaping of Roche equipotentials and reduction of the effective
gravity of the mass-losing star
(Jorissen 2003b; Schuerman 1972; Frankowski \& Tylenda 2001).

\sectionb{4}{THE `TRANSIENT TORUS' SCENARIO}
\label{Sect:scenario}

Gathering the observational and theoretical constraints described above,
we propose a `transient torus' scenario for explaining the observed
orbital periods and eccentricities of ``after AGB'' binaries. This
scenario can be divided into four phases, schematically represented in
Fig.~3:

1. Wind accretion. The system is well detached and the companion accretes
mass and angular momentum from the giant's wind. Spin accretion is
especially
efficient, proceeding through an accretion disk formed around the companion
(Theuns et~al.\ 1996; Mastrodemos \& Morris 1998).
Orbital evolution proceeds roughly as in spherically-symmetric wind case
(Jeans mode), i.e., $a(M_1+M_2) =$ const and the eccentricity stays almost
constant.

2. (Near) RLOF with substantial $L_2/L_3$ outflow. Tidal forces and
evolutionary expansion of the giant bring it closer to its Roche lobe.
The outflow becomes concentrated in the direction to the companion, which
happens even before the actual Roche-lobe filling
(e.g.\ Frankowski \& Tylenda 2001).
The matter is 'funnelled' through the vicinity of $L_1$.

3. Formation of a circumbinary torus. Matter escaping through the
vicinity of $L_2$ (or $L_3$, after mass ratio reversal) forms a spiral
around the system.
But after one orbital period every portion of ejecta becomes shadowed from
the giant by the newly ejected matter and ceases being accelerated outwards
by the radiation pressure on dust. Part of the older ejecta
gravitationally falls back onto the binary and collides with the new stream.
A thick circumbinary torus is formed.

4. Formation of a Keplerian circumbinary disk. The torus drags angular
momentum from the binary and at the same time it is slowly pushed outwards
by the radiation pressure on dust. The leftovers become a Keplerian disk.
Only small part of the ejecta is pulled into Keplerian motion,
so the angular momentum removal from the central binary is moderate
and the orbital period can stay as long as a few hundred days.

Point 2.~in this sequence deserves particular consideration.
At this stage the companion resides within the wind acceleration zone
which is governed by the dust condensation radius, $R_{\rm cond}$.
On TP-AGB $R_{\rm cond}$ is 2--5 $R_{*}$ (Gail \& Sedlmayr 1988).
\begin{wrapfigure}[21]{r}[0pt]{67mm}
\vbox{
\vspace{-4mm}
\centerline{\psfig{figure=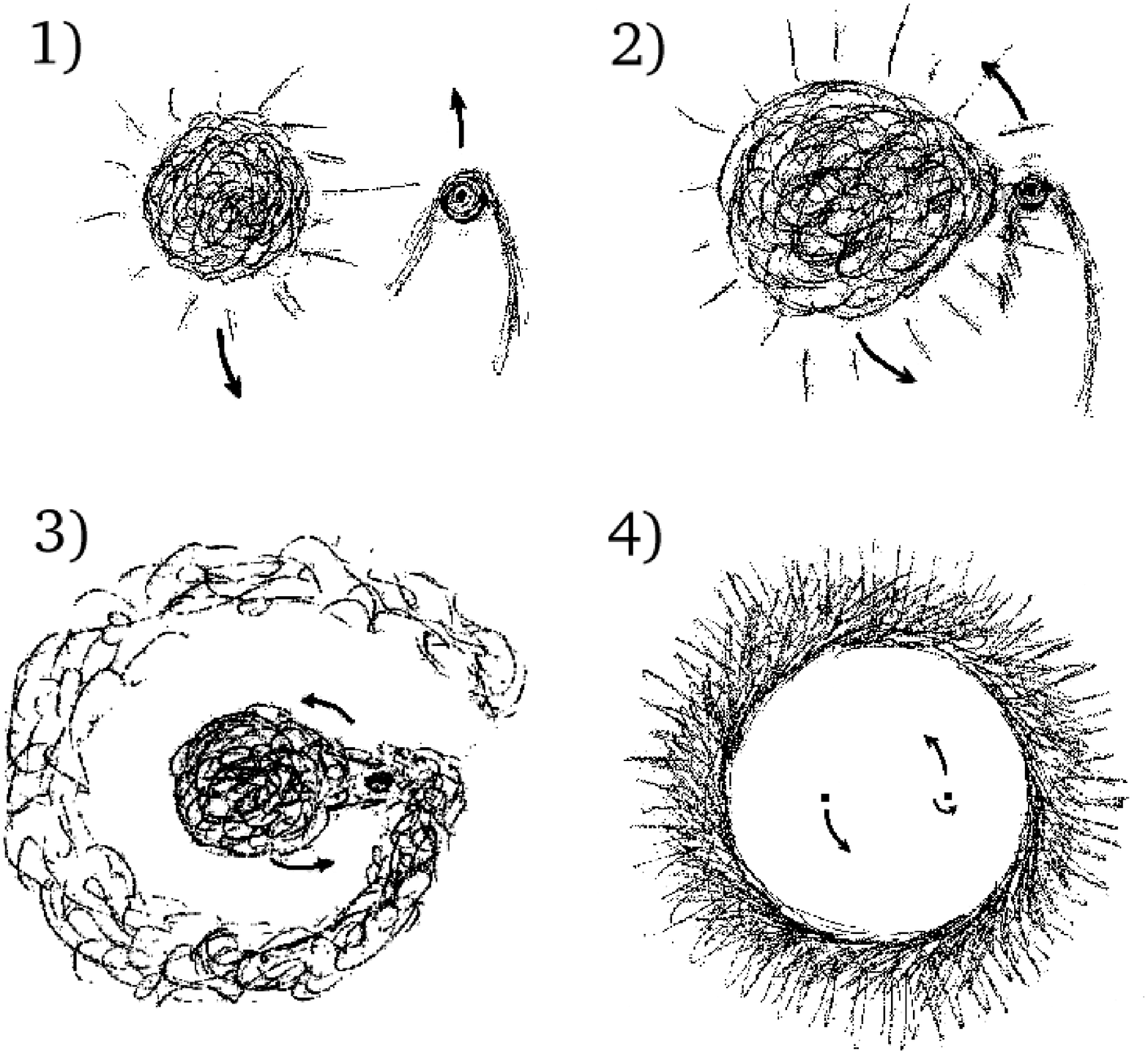,width=74mm,clip=}}
\captionb{3}{\label{Fig:scenario}
The `transient torus' scenario. For description of the
phases, see text.
}
}
\end{wrapfigure}
Thus the matter flowing
preferentially through $L_1$ in the direction of the companion is still
moving slowly in the vicinity of the companion (which favors higher
accretion rate) and is still feeling an
outward acceleration due to radiation pressure on dust (so the
{\em modified} Roche potential is in force and a dynamical mass
{\em transfer} leading to a CE can be avoided).
Mass {\em loss} from the binary can
proceed on a dynamical time scale for some part of this phase without an
ensuing CE.
These effects do not play a role for non-dusty winds, thus not
changing the classical CE at RGB and E-AGB, leading to pre-CV and CV
systems, as required for explaining those close binary populations.

\vskip3mm

\References

\vskip-1mm

\refb
{Artymowicz} P., {Clarke} C.~J., {Lubow} S.~H., {Pringle} J.~E.
1991, ApJ, 370, 35

\refb
{Bond}~H.~E., {Ciardullo}~R., {Meakes}~M.~G.~1993,
in {\it Planetary nebulae}, IAU Symp.~155, eds.~R.~Weinberger
\& A.~Acker, Kluwer Academic Publishers, Dordrecht,~p.~397

\refb
{Clayton} D.~D. 1988, MNRAS, 234, 1

\refb
{Corradi} R., {Schwarz} H.~E. 1997,
in {\it Physical Processes in Symbiotic Binaries and Related
  Systems}, ed. J.~Miko\l ajewska,
  Copernicus Foundation for Polish Astronomy, Warsaw, p. 147

\refb
{Corradi} R.~L.~M., {Brandi} E., {Ferrer} O.~E., {Schwarz} H.~E. 1999, A\&A,
  343, 841

\refb
{Eggleton} P.~P. 1986,
in J. {Tr\"umper}, W.~H.~G. {Lewin}, W. {Brinkmann} (eds.),
  {\it The evolution of galactic X-ray binaries}, Reidel, Dordrecht, p. 87

\refb
{Frankowski} A. 2004,
PhD thesis, N. Copernicus Astronomical Center, Warsaw

\refb
{Frankowski} A., {Jorissen} A. 2006, Obs., 126, 25

\refb
{Frankowski} A., {Tylenda R.} 2001, A\&A, 367, 513

\refb
{Gail} H.-P., {Sedlmayr} E. 1988, A\&A, 206, 153

\refb
{Habing} H. J., Olofsson, H. 2003,
{\it Asymptotic Giant Branch Stars}, Springer Verlag, New York

\refb
{Han} Z., {Podsiadlowski} P., {Eggleton} P.~P. 1994, MNRAS, 270, 121

\refb
{Harpaz} A. 1998, ApJ, 498, 293

\refb
{Iben} I. Jr., {Tutukov} A.~V. 1996, ApJS, 105, 145

\refb
{Jeffries} R.~D., {Stevens} I.~R. 1996, MNRAS, 279, 180

\refb
{Jorissen} A. 2003a,
in R.~L.~M. {Corradi}, J. {Miko\l ajewska}, T.~J. {Mahoney}
(eds.),
  {\it Symbiotic stars probing stellar evolution}, ASP Conf. Ser.,
  303,~25

\refb
{Jorissen} A. 2003b,
in H. {Habing}, H. {Olofsson} (eds.), {\it Asymptotic Giant Branch
Stars},
  Springer Verlag, New York, p.~461

\refb
{Jorissen} A., {Famaey} B., {Dedecker} M., {Pourbaix} D., {Mayor} M., {Udry}
S.
  2004, Rev. Mex. Astron. Astrof. Conf. Ser. 21,  71--72

\refb
{Jorissen} A., {Schmitt} J.~H.~M.~M., {Carquillat} J.~M., {Ginestet} N.,
{Bickert} K.~F. 1996, A\&A, 306, 467

\refb
{Jorissen} A., {Za{\v c}s} L., {Udry} S., {Lindgren} H., {Musaev} F.~A.
2005,
  A\&A, 441, 1135


\refb
{Kenyon} S.~J., {Oliversen} N.~A., {Miko\l ajewska} J., {Mikolajewski} M.,
  {Stencel} R.~E., {Garcia} M.~R., {Anderson} C.~M. 1991, AJ, 101, 637

\refb
{Kenyon} S.~J., {Webbink} R.~F. 1984, ApJ, 279, 252

\refb
{Knapp} G.~R., {Dobrovolsky} S.~I., {Ivezi\'c} Z., {Young} K., {Crosas} M.,
{Mattei} J.~A., {Rupen} M.~P. 1999, A\&A, 351, 97

\refb
{Little} S.~J., {Little-Marenin} I.~R., {Bauer} W.~H. 1987, AJ, 94, 981

\refb
{Masseron} T., {Van Eck} S., {Famaey} B., {Goriely} S., {Plez} B., {Siess}
L.,
  {Beers} T., {Primas} F., {Jorissen} A. 2006, A\&A, 455, 1059

\refb
{Mastrodemos} N., {Morris} M. 1998, ApJ, 497, 303

\refb
{McClure} R.~D., {Fletcher} J.~M., {Nemec} J.~M. 1980, ApJ, 238, 35

\refb
{Miko\l ajewska} J. 2003,
in R.~L.~M. {Corradi}, J. {Miko\l ajewska}, T.~J. {Mahoney}
(eds.),
  {\it Symbiotic stars probing stellar evolution}, ASP Conf. Ser.,
  303,~9

\refb
{Miko\l ajewska} J., {Kenyon} S.~J. 1992, MNRAS,  177

\refb
{Nelemans} G., {Verbunt} F., {Yungelson} L.~R., {Portegies Zwart} S.~F.
2000, A\&A, 1011

\refb
{Pols} O. R., {Karakas} A. I., {Lattanzio} J. C., {Tout} C. A. 2003
in R.~L.~M. {Corradi}, J. {Miko\l ajewska}, T.~J. {Mahoney}
(eds.),
  {\it Symbiotic stars probing stellar evolution}, ASP Conf. Ser.,
  303, 290

\refb
{Quiroga} C., {Miko{\l}ajewska} J., {Brandi} E., {Ferrer} O.,
{Garc{\'{\i}}a}
  L. 2002, A\&A, 387, 139

\refb
{Sahai} R. 1992, A\&A, 253, 33

\refb
{Schild} H., {Boyle} S.~J., {Schmid} H.~M. 1992, MNRAS, 258, 95

\refb
{Schmid} H.~M., {Nussbaumer} H. 1993, A\&A, 268, 159

\refb
{Schmidt} M., {Miko\l ajewska} J. 2003,
in R. {Corradi}, J. {Miko\l ajewska}, T.~J. {Mahoney} (eds.),
  {\it Symbiotic stars probing stellar evolution}, ASP Conf. Ser.,
  303, 163

\refb
{Schuerman} D.~W. 1972, Ap\&SS, 19, 351

\refb
{Schwarz} H.~E. 1991, A\&A, 243, 469

\refb
{Soker} N. 2000, A\&A, 357, 557

\refb
{Theuns} T., {Boffin} H. M.~J., {Jorissen} A. 1996, MNRAS, 280, 1264

\refb
{Th\'evenin} F., {Jasniewicz} G. 1997, A\&A, 320, 913

\refb
{Van Eck} S., {Goriely} S., {Jorissen} A., {Plez} B. 2003, A\&A, 404, 291

\refb
{Van Winckel} H. 2003, ARA\&A, 41, 391

\refb
{Waelkens} C., {Van Winckel} H., {Waters} L.~B.~F.~M., {Bakker} E.~J.
1996, A\&A,~314,~17

\end{document}